\documentclass[final]{ws-mpla}

\usepackage{epsfig}
\usepackage{color}
\usepackage{colortab}
\usepackage{array}
\usepackage{rotating}
\usepackage{hyperref}

\newcommand{\beq}{\begin{equation}}
\newcommand{\eeq}{\end{equation}}
\newcommand{\bdis}{\begin{displaymath}}
\newcommand{\edis}{\end{displaymath}}
\newcommand{\bea}{\begin{eqnarray}}
\newcommand{\eea}{\end{eqnarray}}
\newcommand{\barr}{\begin{array}}
\newcommand{\earr}{\end{array}}

\newcommand{\X}{{\cal X}}
{}

\newcommand{\eg}{e.g.}
\newcommand{\ie}{{\it i.e}.}

\def\be{\begin{equation}}
\def\ee{\end{equation}}
\def\bea{\begin{eqnarray}}
\def\eea{\end{eqnarray}}

\begin{document}
\markboth{R. Faccini, A. Pilloni, A. Polosa}
{Exotic Heavy Quarkonium Spectroscopy: A Mini-review}

\catchline{}{}{}{}{}

\title{Exotic Heavy Quarkonium Spectroscopy: A Minireview}

\author{\footnotesize RICCARDO FACCINI\footnote{riccardo.faccini@roma1.infn.it}, ALESSANDRO PILLONI, ANTONIO D. POLOSA}

\address{Sapienza University of Roma and INFN Roma\\
Dipartimento di Fisica, Roma, P.le A. Moro 5, 00.185, Roma
Italy}

\maketitle

\pub{Received 6 July 2012}{Accepted 6 July 2012}

\begin{abstract}
Since nine years experiments have been observing a host of  exotic states decaying into heavy quarkonia. The interpretation of most of them still remains  uncertain and, in some cases, controversial, notwithstanding a considerable progress has been made on the quality of the experimental information available and a number of ideas and models have been put forward to explain the observations.

In this mini-review we will summarize the measurements, with the most recent updates, and 
list the useful ones yet to be done. We will discuss the problem of the spin of the $X$, which could hide some major surprise on its interpretation, and review some more phenomenological issues debated in the field.   

\keywords{spectroscopy, exotic states, heavy quarks}
\end{abstract}

\ccode{PACS Nos.: 14.40.Rt, 14.40.Pq}

\section{General Overview}
Since 2003 the experimental study of  charmonium-like $X,Y,Z$ resonances  has produced a remarkable amount of information about their masses, widths and $J^{PC}$ quantum numbers.
There are several hints that most of these states cannot fit standard charmonium interpretations. 
The strength of such statements is based on the solid knowledge of quarkonia level structure and decay patterns both in charm and beauty sectors.

Nevertheless, the bulk of present experimental data has not  proven to be sufficient to  identify
clear patterns pointing at a unified description of those $X,Y,Z$ states  having most likely an exotic structure.
Studying their features offers the concrete opportunity to discover the new strong interaction dynamics  which is  suspected to be at work in their production and quark structure. To aim at this objective the continuity in the experimental investigation should be guaranteed.    

The  possibility of hadron structures different from mesons and baryons  
has been left open since the early days of the quark model~\cite{GellMann:1964nj}. 
In  later studies it has  often been observed  that multiquark hadrons could  be extremely broad states,  escaping the experimental identification (see e.g. Ref.~\cite{witten}). A turning point has been reached in the pentaquark days when a narrow structure with two diquarks and a quark was proposed in Ref.~\cite{jaffe}. Following that, diquark-antidiquark structures were suggested to explain the decay patterns of light scalar mesons~\cite{mainoi1,mainoi2} and heavy-light diquarks were introduced to study the $X,Y,Z$ spectroscopy~\cite{Maiani:2004vq}.  

Another option is that some of these resonances could be hadron molecules having a mass almost exactly  at the threshold value for  the fall apart dissociation in their constituents. The prototypical example of this phenomenon is assumed to be displayed by the $X(3872)$ modeled as a large size $D\bar D^{*}$ molecule.  Notwithstanding the fact that this picture is at odds with the simplest cross section studies of $X$ prompt production at hadron colliders~\cite{Bignamini:2009sk,Bignamini:2009fn}, there is ongoing research on how to explain this phenomenon in terms of Feshbach molecules or resorting to final-state-interactions, as made in Ref.~\cite{Artoisenet:2009wk}.   
The molecular option would unveil a subtle  behavior of strong interactions on length scales as large as  $10$~fm.

The actual identification of $X,Y,Z$ states
would represent a major progress in our understanding of elementary particles and it might  imply the prediction of a large number of additional states that have not yet been observed.

The most likely possible states beyond  mesons and 
baryons are:

\begin{itemize}

\item {\bf Hybrids:} bound states of a quark-antiquark pair and a number of
constituent gluons. The lowest-lying state is expected to have quantum numbers
$J^{PC}=0^{+-}$. Since a quarkonium state cannot have
these quantum numbers, this is a unique signature for
hybrids. An additional signature is the preference for a hybrid to
decay into quarkonium and a state that can be produced by the excited
gluons (\eg\ $\pi^+\pi^-$ pairs); see \eg~Ref.~\cite{ibridi}.
\item {\bf Molecules:} bound states of two mesons, usually represented as
$[Q\bar{q}][q^{\prime}\bar{Q}]$, where $Q$ is the heavy quark. The
system would be stable if the binding energy were to set the mass of the
states below the sum of the two meson masses.
The bound state could decay into its constituents~\cite{Braaten:2003he}, \eg,~the $X(3872)$, supposedly a $D\bar D^*$ molecule, could dissociate into $DD\pi$. For this reason the width of the $X(3872)$ is expected to be as large as the width of the $D^*$, which is about $100$~keV.
Present measurements of $\Gamma_X$ find $\Gamma_X\lesssim 1.7$~MeV. A value of $\Gamma_X$ sensibly larger that $100$~keV would allow little space to the molecule option.\item {\bf Tetraquarks:} a  quark pair (diquark), neutralizing its color with an
antiquark pair, usually represented as $[Qq][\bar{q^{\prime}}\bar{Q}]$. A full
nonet of states is predicted for each spin-parity, \ie, a large number
of states are expected. There is no need for these states to be close to
any threshold~\cite{Maiani:2004vq}.

\end{itemize}

A way of `hybridizing' between molecules and  compact multiquarks  is proposed in the hadrocharmonium model where a $c\bar c $ core is supposed to be surrounded by a light quark cloud as described in~\cite{Dubynskiy:2008mq}. In addition to this there is the
lurking possibility that some of the observed states are misinterpretations of
threshold effects: a given amplitude might be enhanced when new hadronic final
 states become energetically possible, even in the absence of resonances.

\section{Experimental overview}
Exotic heavy quarkonia candidates have been observed both at $e^+e^-$ and hadron machines.
Besides finding the states, an attempt was made 
to measure their quantum numbers, in particular their spin and charge and parity eigenvalues, $J^{PC}$. To achieve this, different production and decay modes were considered.

There are several possible production mechanisms of exotic particles at $e^+e^-$ machines. Let us generically call them  ${\cal X}$. 
\begin{itemize}
\item In association with Initial State Radiation (ISR), \ie, $e^+e^-\to \X \gamma_{ISR}$. Such production is possible only if $J^{PC}=1^{--}$
\item In the fusion of two virtual photons ($\gamma\gamma$ production). Such process is allowed only if $C=+$
\item In conjunction with a charmonium state (e.g. $e^+e^-\to J/\psi \X$). In this case only states with a value $C$ opposite to the one of the associated charmonium  can be produced  
\item In $B$ decays, typically in association with a kaon, \ie, $B\to \X K$. This mechanism is allowed for any $J^{PC}$, albeit preferred for low values of the spin
\item From direct production $e^+e^-\to \X$. This is feasible only for states with $J^{PC}=1^{--}$ and if the center of mass energy of the machine coincides with the mass of the exotic state
\item In conjunction with a light meson, typically a pion, {\it i.e.},  $e^+e^-\to \X\pi$, with no restriction on $J^{PC}=1^{--}$. For this process to occur with a reasonable probability the
mass of the $\X$ state must not differ  more than a pion mass from the center of mass energy 
\end{itemize}
The first four production mechanisms are specific of states with charm content, while in the past generation of $B-$Factories only states with bottom content could be produced with the last two.

As far as decays are concerned, all final states are accessible at $e^+e^-$  experiments, although states with $D$ mesons are systematically disfavored by the fact that the observable final states have small branching fractions. 

Hadron colliders instead produce quarkonium states either directly or in $B$ decays. In any case the search is typically conducted inclusively, \ie, without distinguishing the production mechanism and  final states with neutral particles are extremely hard to study. Regardless of the more limited range of channels that can be studied, hadron colliders are very important in this research field both because of their high statistics and because the production at hadron colliders can by itself provide a lot of information on the nature of the states (see for instance Ref.~\cite{Bignamini:2009sk}).

\subsection{Charmonium-like states}
\begin{figure}[t]
\begin{center}
\epsfig{file=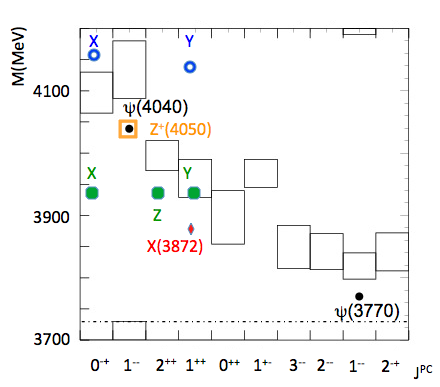,height=5.3cm}
\epsfig{file=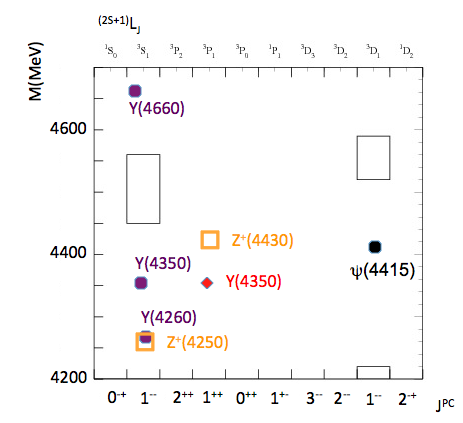,height=5.9cm}
\caption{\it Observed states with hidden charm above the open charm threshold catalogued according to the most likely $J^{PC}$ assignment. The theory predictions for the masses of the regular charmonium levels are given according to the potential models 
described in Ref.~\protect\cite{Brambilla:2004wf}. }
\label{fig:charmoniumSummary}
\end{center}
\end{figure}

The large number of states with  charmonium content is summarized in 
Fig~\ref{fig:charmoniumSummary}\footnote{see Ref.~\cite{Drenska:2010kg} for a comprehensive review}.
The status of the experimental observations is extremely fragmented. Pictorial summaries, separated by production mechanisms, are reported in the updated  Tabs.~\ref{tab:obsmatrB}-\ref{tab:obsmatrCh}.
The exotic states have been observed always in only one production mechanism and often in only one final state. None of the states has been searched systematically in all final states. Sometimes the analysis of a given final state for a given production mechanism is missing, but often it has either been performed only in a limited mass range or the invariant mass spectrum has been published without a fit to the possible new state. This is mostly due to the fact that some of the analyses did not show a significant signal themselves and they were published before a new state was observed. 

Another remarkable aspect is that, given the limited statistics, the quantum number assignment, when not forced by the production mechanism, is extremely uncertain. Even the best known state, the $X(3872)$, has two possible quantum number assignments $J^{PC}=1^{++}$ or $2^{-+}$, as it will be detailed in Sec.~\ref{sec:spin}. 

\begin{table}[!htb]
\tbl{\it Status of the searches of the new states  in the process $B\to X K$, $X\to f$, for several final states $f$. 
Final states where each exotic states were observed (``S") or excluded (``N/S") are indicated. A final states is marked as ``N" if the analysis has not been performed in a given mass range and with ``M/F" if the spectra are published but a fit to a given state has not been performed. Finally ``N/A" indicates that  quantum numbers forbid the decay and ``N/F" if an analysis is experimentally too challenging. (Default)}
{
\begin{tiny}
\begin{tabular}{| l ||@{}c@{}|@{}c@{}|@{}c@{}|@{}c@{}|@{}c@{}|@{}c@{}|@{}c@{}|@{}c@{}|@{}c@{}|@{}c@{}|@{}c@{}|@{}c@{}|@{}c@{}|@{}c@{}|@{}c@{}|@{}c@{}|@{}c@{}|@{}c@{}|} \hline
 & $J^{PC}$ &  $J/\psi\pi\pi$ &  $J/\psi\omega$ &  $J/\psi\gamma$ &  $J/\psi\phi$ &  $J/\psi\eta$ & $\psi(\rm{2S})\pi\pi$ & $\psi(\rm{2S})\omega$ & $\psi({\rm 2S})\gamma$ & $\chi_c\gamma$ & pp & $\Lambda\Lambda$ & $\Lambda_c\Lambda_c$ & $DD$ &$ DD^*$ & $D^*D^*$ & $D_s^{(*)}D_s^{(*)}$ & $\gamma\gamma$\\\hline\hline
$X(3872)$ & $1^{++} $&{\green\bf S}&{ \green\bf S}&{ \green\bf S}& N/A & { \red\bf N/S} & N/A & N/A &{ \green\bf S}& { \red\bf N/S} & { \cyan\bf M/F} & { \cyan\bf M/F} & N/A & N/A &{ \green\bf S}& N/A & N/A & { \red\bf N/S}\\ 
& $2^{-+}$ &&&&&&&&&&&&&&&&&\\ \hline
$X(3940)$ & $J^{P+}$& { \cyan\bf M/F} &{ \green\bf S}& { \red\bf N/S} & N/A & N/A & N/A & N/A & { \cyan\bf M/F} & N/A & { \cyan\bf M/F} & { \cyan\bf M/F} & N/A & { \cyan\bf M/F} & { \red\bf N/S} & N/A &{\cyan \bf N}&{\cyan \bf N}\\ \hline
$Z(3940)$ & $2^{++}$ & { \cyan\bf M/F} & { \cyan\bf M/F} & { \red\bf N/S} & N/A & N/A & N/A & N/A & { \cyan\bf M/F} & N/A & { \cyan\bf M/F} & { \cyan\bf M/F} & N/A & { \cyan\bf M/F} & { \cyan\bf M/F} & N/A &{\cyan \bf N}&{\cyan \bf N}\\ \hline
$Y(4140)$ & $J^{P+}$ & { \cyan\bf M/F} & { \cyan\bf M/F} &{\cyan \bf N}&{ \green\bf S}& N/A &{\cyan \bf N}& N/A &{\cyan \bf N}& N/A & { \cyan\bf M/F} & { \cyan\bf M/F} & N/A & { \cyan\bf M/F} &{\cyan \bf N}&{\cyan \bf N}&{\cyan \bf N}&{\cyan \bf N}\\ \hline
$X(4160)$ & $0^{P+}$ & { \cyan\bf M/F} & { \cyan\bf M/F} &{\cyan \bf N}& { \cyan\bf M/F} & N/A &{\cyan \bf N}& N/A &{\cyan \bf N}& N/A & { \cyan\bf M/F} & { \cyan\bf M/F} & N/A & { \cyan\bf M/F} &{\cyan \bf N}&{\cyan \bf N}&{\cyan \bf N}&{\cyan \bf N}\\ \hline
$Y(4260)$ & $1^{--}$ &{ \green\bf S}& N/A & N/A & N/A & { \cyan\bf M/F} &{\cyan \bf N}& N/A & N/A &{\cyan \bf N}& { \cyan\bf M/F} & { \cyan\bf M/F} & N/A &{\cyan \bf N}&{\cyan \bf N}&{\cyan \bf N}&{\cyan \bf N}& N/A\\ \hline
$X(4350)$ & $J^{P+}$ & { \cyan\bf M/F} & { \cyan\bf M/F} &{\cyan \bf N}& { \cyan\bf M/F} & N/A &{\cyan \bf N}&{\cyan \bf N}&{\cyan \bf N}& N/A & { \cyan\bf M/F} & { \cyan\bf M/F} & N/A &{\cyan \bf N}&{\cyan \bf N}&{\cyan \bf N}&{\cyan \bf N}&{\cyan \bf N}\\ \hline
$Y(4350)$ & $1^{--}$ & { \cyan\bf M/F} & N/A & N/A & N/A & { \cyan\bf M/F} &{\cyan \bf N}& N/A & N/A &{\cyan \bf N}& { \cyan\bf M/F} & { \cyan\bf M/F} & N/A &{\cyan \bf N}&{\cyan \bf N}&{\cyan \bf N}&{\cyan \bf N}& N/A\\ \hline
$Y(4660)$ & $1^{--}$ &{\cyan \bf N}& N/A & N/A & N/A  & { \cyan\bf M/F} &{\cyan \bf N}& N/A & N/A &{\cyan \bf N}& { \cyan\bf M/F} & { \cyan\bf M/F} & { \cyan\bf M/F} &{\cyan \bf N}&{\cyan \bf N}&{\cyan \bf N}&{\cyan \bf N}& N/A\\ \hline
\end{tabular}
\end{tiny}
}
\label{tab:obsmatrB}
\end{table}

\begin{table}[!htb]
\tbl{\it Status of the searches of the new states  in the process $e^+e^-\to X\gamma_{ISR}$, $X\to f$, for several final states $f$. The meaning of the symbols is explained in the caption of Tab.~\ref{tab:obsmatrB}}
{
\begin{tiny}
\begin{tabular}{| l ||@{}c@{}|@{}c@{}|@{}c@{}|@{}c@{}|@{}c@{}|@{}c@{}|@{}c@{}|@{}c@{}|@{}c@{}|@{}c@{}|@{}c@{}|@{}c@{}|} \hline
&$J^{PC}$ &  $J/\psi\pi\pi$ &   $\psi(\rm{2S})\pi\pi$ & $J\psi\eta$ &  $\chi_c\gamma$ & pp & $\Lambda\Lambda$ & $\Lambda_c\Lambda_c$ & $DD$ &$ DD^*$ & $D^*D^*$ & $D_s^{(*)}D_s^{(*)}$ \\ \hline \hline
$Y(4260)$ & $1^{--}$ & { \green \bf S}&  { \red\bf N/S} &  { \red\bf N/S} &  { \red\bf N/S} &  { \red\bf N/S} & { \cyan\bf M/F} & N/A &  { \red\bf N/S} &  { \red\bf N/S} &  { \red\bf N/S} &  { \red\bf N/S}\\ \hline
$Y(4350)$ & $1^{--}$ &  { \red\bf N/S} & { \green \bf S}& { \cyan\bf M/F} & { \cyan\bf M/F} & { \cyan\bf M/F} & { \cyan\bf M/F} & N/A & { \cyan\bf M/F} & { \cyan\bf M/F} & { \cyan\bf M/F} & { \cyan\bf M/F}\\ \hline
$Y(4660)$ & $1^{--}$ &  { \red\bf N/S} & { \green \bf S}& { \cyan\bf M/F} & { \cyan\bf M/F} & { \cyan\bf M/F} & { \cyan\bf M/F} & { \green \bf S}& { \cyan\bf M/F} & { \cyan\bf M/F} & { \cyan\bf M/F} & { \cyan\bf M/F}\\ \hline
\end{tabular}
\end{tiny}
}
\label{tab:obsmatrISR}
\end{table}

Looking into this ``observational" tables in detail, $B$ decays (Tab.~\ref{tab:obsmatrB}) are the most studied, but there is a significant amount of missing fits (``M/F"). Particularly severe is the lack of analysis of the baryonic spectra especially in consideration that  baryonic decays are a signature of tetraquark states. Some other modes have never been studied, mostly because the number of expected events is very low. Nonetheless the study of $B\to \psi(2S)\pi\pi K$ decays should be relatively clean, while $D^*\bar{D}^*$ and above all $D_s^{(*)}\bar{D}_s^{(*)}$ suffer from the low branching fractions of the observed states.

Resonances produced in conjunction with an initial state radiation (ISR) photon  (Tab.~\ref{tab:obsmatrISR})  have an unambiguous $J^{PC}=1^{--}$ assignment and therefore fewer analyses are needed to establish their properties. Nonetheless it is striking to see that a large fraction of analyses have been carried out exclusively for the first observed exotic state, the $Y(4260)$. It can also be noticed that no search is published involving $D_s^{(*)}$ mesons: while the efficiency is expected to be very low, background should be low as well and surprises can  arise. Finally the $Y(4660)$ has been object of one of the combined analyses we are advocating here~\cite{Cotugno:2009ys}: two resonances apparently different, observed in $\psi(2S)$ and $\Lambda_c{\bar{\Lambda}_c}$ final states, if fitted under the same ansatz were found to be consistent with being the same and interesting ratios of branching fractions were measured.

\begin{table}[!htb]
\tbl{\it Status of the searches of the new states  in the process $e^+e^-\to X J/\psi$, $X\to f$, for several final states $f$. The meaning of the symbols is explained in the caption of Tab.~\ref{tab:obsmatrB}}
{
\begin{tiny}
\begin{tabular}{| l ||@{}c@{}|@{}c@{}|@{}c@{}|@{}c@{}|@{}c@{}|@{}c@{}|@{}c@{}|@{}c@{}|@{}c@{}|@{}c@{}|@{}c@{}|@{}c@{}|@{}c@{}|@{}c@{}|@{}c@{}|@{}c@{}|} \hline
 & $J^{PC}$ &  $J/\psi\pi\pi$ &  $J/\psi\omega$ &  $J/\psi\gamma$ &  $J/\psi\phi$ & $\psi(\rm{2S})\pi\pi$ & $\psi(\rm{2S})\omega$ & $\psi({\rm 2S})\gamma$ & $\chi_c\gamma$ & pp & $\Lambda\Lambda$ & $\Lambda_c\Lambda_c$ & $DD$ &$ DD^*$ & $D^*D^*$\\\hline\hline
$X(3872)$ & $1^{++} $[$2^{-+}$] & N/F &{\cyan \bf N}& N/F & N/A & N/F & N/A & N/F & N/F & N/F & N/F & N/A &{ \cyan\bf M/F} &{ \cyan\bf M/F} & N/A\\\hline
``$X / Y (3940)$'' & $0^{-+}$ [$J^{P+}$] & N/F &{\cyan \bf N}& N/F & N/A & N/F & N/A & N/F & N/F & N/F & N/F & N/A &{ \green\bf S} &{ \cyan\bf M/F} & N/A\\\hline
$Z(3940)$ & $2^{++} $& N/F &{\cyan \bf N}& N/F & N/A & N/F & N/A & N/F & N/F & N/F & N/F & N/A &{ \cyan\bf M/F} &{ \cyan\bf M/F} & N/A\\\hline
$Y(4140)$ & $J^{P+}$ & N/F &{\cyan \bf N}& N/F &{\cyan \bf N}& N/F & N/A & N/F & N/F & N/F & N/F & N/A &{ \cyan\bf M/F} &{ \cyan\bf M/F} &{ \cyan\bf M/F}\\\hline
$X(4160)$ & $0^{P+} $& N/F &{\cyan \bf N}& N/F &{\cyan \bf N}& N/F & N/A & N/F & N/F & N/F & N/F & N/A &{ \cyan\bf M/F} &{ \green\bf S}&{ \cyan\bf M/F}\\\hline
$X(4350)$ & $J^{P+}$ & N/F &{\cyan \bf N}& N/F &{\cyan \bf N}& N/F &{\cyan \bf N}& N/F & N/F & N/F & N/F & N/F &{ \cyan\bf M/F} &{ \cyan\bf M/F} &{ \cyan\bf M/F}\\\hline
\end{tabular}
\end{tiny}
}
\label{tab:obsmatrRec}
\end{table}

\begin{table}[!htb]
\tbl{\it Status of the searches of the new states  in the process  $\gamma \gamma\to X$, $X\to f$, for several final states $f$. The meaning of the symbols is explained in the caption of Tab.~\ref{tab:obsmatrB}}
{
\begin{tiny}
\begin{tabular}{| l ||@{}c@{}|@{}c@{}|@{}c@{}|@{}c@{}|@{}c@{}|@{}c@{}|@{}c@{}|@{}c@{}|@{}c@{}|@{}c@{}|@{}c@{}|@{}c@{}|@{}c@{}|@{}c@{}|@{}c@{}|@{}c@{}|@{}c@{}|} \hline
 & $J^{PC}$ &  $J/\psi\pi\pi$ &  $J/\psi\omega$ &  $J/\psi\gamma$ &  $J/\psi\phi$ & $\psi(\rm{2S})\pi\pi$ & $\psi(\rm{2S})\omega$ & $\psi({\rm 2S})\gamma$ &  pp & $\Lambda\Lambda$ & $\Lambda_c\Lambda_c$ & $DD$ &$ DD^*$ & $D^*D^*$ & $D_s^{(*)}D_s^{(*)}$\\\hline\hline
$X(3872)$ &$ 1^{++}$ [$2^{-+}$] &{\cyan \bf N}& N/F & N/F & N/A & N/A & N/A & N/F & {\cyan \bf M/F} & {\cyan \bf M/F} & N/A & {\cyan \bf M/F} &{\cyan \bf N}& N/A & N/A\\\hline
``$X / Y (3940)$'' & $0^{-+}$ [$J^{P+}$] &{\cyan \bf N}&{\green \bf S}& N/F & N/A & N/A & N/A & N/F & {\cyan \bf M/F} & {\cyan \bf M/F} & N/A & {\green \bf S} &{\cyan \bf N}& N/A & N\\\hline
$Z(3940)$ & $2^{++}$ &{\cyan \bf N}&{\green \bf S}   & N/F & N/A & N/A & N/A & N/F & {\cyan \bf M/F} & {\cyan \bf M/F} & N/A &{\green \bf S}&{\cyan \bf N}& N/A & N\\\hline
$Y(4140)$ & $J^{P+}$ &{\cyan \bf N}& {\cyan \bf M/F} & N/F &{\red \bf N/S}&{\cyan \bf N}& N/A & N/F &{\cyan \bf N}&{\cyan \bf N}& N/A & {\cyan \bf M/F} &{\cyan \bf N}&{\cyan \bf N}& N\\\hline
$X(4160)$ & $0^{P+}$ &{\cyan \bf N}& {\cyan \bf M/F} & N/F &{\red \bf N/S}&{\cyan \bf N}& N/A & N/F &{\cyan \bf N}&{\cyan \bf N}& N/A & {\cyan \bf M/F} &{\cyan \bf N}&{\cyan \bf N}& N\\\hline
$X(4350)$ & $J^{P+}$ &{\cyan \bf N}&{\cyan \bf N}& N/F &{\green \bf S}&{\cyan \bf N}&{\cyan \bf N}& N/F &{\cyan \bf N}&{\cyan \bf N}&{\cyan \bf N}&{\cyan \bf N}&{\cyan \bf N}&{\cyan \bf N}& N\\\hline
\end{tabular}
\end{tiny}
}
\label{tab:obsmatrGG}
\end{table}

On the recoil of a $J/\psi$ and in $\gamma\gamma$ interactions  (Tabs.~\ref{tab:obsmatrRec} and ~\ref{tab:obsmatrGG})  only $C=+$ neutral states can be observed. This restricts the number of final states of interest. Also, the low multiplicity of these decays and the large missing momentum in the case of $\gamma\gamma$ decays makes these analyses experimentally challenging. On the other side, $C=+$ states are the least known and reinforcing the evidence of the signals would help.
It is also interesting to notice that, mostly due to statistics, the recoil to any other particle but the $J/\psi$ has not been investigated. Given the selection rules the recoil to $\chi_{c0}$ or $\chi_{c2}$ would be very interesting. 

 Concerning the searches of charged exotic resonances, the most striking signature of states made of more than two quarks, very few  have
  been conducted in $B$ decays, as shown in Tab.~\ref{tab:obsmatrCh}. While we believe that for each exotic neutral spectrum the corresponding charged particle should be searched, only five combinations of final states and exotic states has been searched for. As an example, no information has been extracted from Refs.~\cite{:2007wg,Mizuk:2008me} on the charged partner of the $X(3872)$, $Z(3870)$ in our table, which has long been pointed out as a critical state to search for.  
  \begin{table}[!htb]
\tbl{\it Status of the searches of the new charged states in several final states. The meaning of the symbols is explained in the caption of Tab.~\ref{tab:obsmatrB}}
{
\begin{tiny}
  \begin{tabular}{| l ||@{}c@{}|@{}c@{}|@{}c@{}|@{}c@{}|@{}c@{}|@{}c@{}|@{}c@{}|@{}c@{}|} \hline
 &$J/\psi\pi$&$J/\psi\pi\pi^0$&$\psi(2S)\pi$&$\psi(2S)\pi\pi^0$&$\chi_{c1}\pi$&$DD$&$DD^*$&$D^*D^*$\\ \hline\hline
$Z^+(3870)$& {\cyan \bf M/F} &N/S& {\cyan \bf M/F} &{\cyan \bf N}& {\cyan \bf M/F} &{\cyan \bf N}&{\cyan \bf N}&N/A\\ \hline
$Z^+(3940)$& {\cyan \bf M/F} &{\cyan \bf N}& {\cyan \bf M/F} &{\cyan \bf N}& {\cyan \bf M/F} &{\cyan \bf N}&{\cyan \bf N}&N/A\\ \hline
$Z^+(4050)$& {\cyan \bf M/F} &{\cyan \bf N}& {\cyan \bf M/F} &{\cyan \bf N}&{\green \bf S}   &{\cyan \bf N}&{\cyan \bf N}& {\cyan \bf M/F} \\ \hline
$Z^+(4140)$& {\cyan \bf M/F} &{\cyan \bf N}& {\cyan \bf M/F} &{\cyan \bf N}& {\cyan \bf M/F} &{\cyan \bf N}&{\cyan \bf N}& {\cyan \bf M/F} \\ \hline
$Z^+(4250)$& {\cyan \bf M/F} &{\cyan \bf N}& {\cyan \bf M/F} &{\cyan \bf N}&{\green \bf S}   &{\cyan \bf N}&{\cyan \bf N}& {\cyan \bf M/F} \\ \hline
$Z^+(4350)$& {\cyan \bf M/F} &{\cyan \bf N}& {\cyan \bf M/F} &{\cyan \bf N}& {\cyan \bf M/F} &{\cyan \bf N}&{\cyan \bf N}& {\cyan \bf M/F} \\ \hline
$Z^+(4430)$&N/S&{\cyan \bf N}&{\green \bf S}   &{\cyan \bf N}& {\cyan \bf M/F} &{\cyan \bf N}&{\cyan \bf N}& {\cyan \bf M/F} \\ \hline
$Z^+(4660)$& {\cyan \bf M/F} &{\cyan \bf N}& {\cyan \bf M/F} &{\cyan \bf N}& {\cyan \bf M/F} &{\cyan \bf N}&{\cyan \bf N}& {\cyan \bf M/F} \\ \hline 
\end{tabular}
\end{tiny}
}
\label{tab:obsmatrCh}
\end{table}

Searches at hadron colliders  have the advantage to have very large samples and this can solve some of the existing puzzles, like the one on the charged states that can be looked into the $J/\psi\pi^+$ spectrum. On the other side, backgrounds are very high and therefore final states with too high multiplicity and above all with neutral particles are not at reach.  Finally, it is hard to extract information on production cross sections and therefore branching fractions. Nonetheless, it is clear that a systematic search in $p\bar{p}$ collisions is likely to clarify the picture significantly. For instance there is a tension between the LHCb~\cite{Aaij:2012pz} and CDF~\cite{Aaltonen:2009tz} searches for a $X(4140)$ state produced in $B$ decays and decaying into $J/\psi \phi$. Critical information about such state could come from the search of its production  prompt $p\bar{p}$ interactions.
 
 \subsection{Bottomonium-like states}

Exotic particles with bottom content can be searched either at hadron colliders or at $e^+e^-$ machines with energy scans above the $\Upsilon(4S)$. The search for neutral such particles in direct production ($e^+e^-\to Y_b$) returned no evidence~\cite{belle_Y5sYpipi,belle_scan,babar_scan}. The search potential of $B$-Factories is limited by the fact that bottomonium at masses higher than the open bottom threshold can only be produced of $J^{PC}=1^{--}$ due to the restricted phase space,
and that  the accessible mass range is full of threshold openings. 

The most striking observation in the field came instead from the search of a charged state with bottomonium content (the $Z_b$) : two such states where observed by Belle in $\Upsilon(5\rm S)\to Z_b\pi$~\cite{Belle:2011aa}, when running at center-of-mass energies above the $\Upsilon(4\rm S)$. This is a clear evidence of states with four constituent quarks and it would be important to confirm it, but the statistics collected by BaBar at those energies is not enough.

\section{Open Issues}
We will focus mainly on one open question: which are the quantum number of the $X$? The possibility that the $1^{++}$ assignment were wrong potentially hides surprises in its constitution in terms of quarks or hadrons.  Some qualitative considerations about the molecule/tetraquark discussion will be added. 

\subsection{The X spin problem}
\label{sec:spin}
As mentioned above, notwithstanding the $X$ particle is the most studied one and was the first to be discovered, its $J^{PC}$ quantum numbers have not yet been established.
Following the analysis by CDF~\cite{Abulencia:2005zc}, two options are possible for its quantum numbers, either $1^{++}$ or $2^{-+}$.

The fact that it might be a $2^{-+}$ state is rather challenging for its interpretation: $i)$ the loosely bound molecule mechanism is excluded 
$ii)$ it could be a $D-$wave, $1^1D_2$, charmonium $iii)$ it could still be a tetraquark but a unit of orbital angular momentum between the diquarks has to be added to reach the overall odd parity. Each diquark is supposed to be in its $\ell = 0$ configuration ($0^-$ and $1^-$ diquarks such as $\bar q_c q$ and $\bar q_c \boldsymbol \gamma \gamma_5q$
are instead assumed to be zero in the `single mode' approximation~\cite{Jaffe:2004ph}). 
Nevertheless such an assignment would call for more states lower than the X.

Let us focus on the charmonium option $ii)$. We have investigated it in~\cite{Burns:2010qq} using an hadron string picture in the heavy quark limit (quark masses at the ends of the string much higher that the string tension). This simple model proved to be rather efficient at computing the masses of orbitally excited quarkonium states in the $b-$quark sector even better than in the charm one, as expected by the infinite quark mass limit. The $3872$~MeV mass is found to be $\sim 100$~MeV heavier than the one computed for a $1^1D_2$ state whereas all other standard states ($P,D$-wave bottomonia, $P$-wave charmonia) are computed with a precision ranging between $1-10$~MeV with the exception of another charm $D$-wave, the $\psi(3770)$, which is found to be $\sim 50$~MeV heavier than the predicted $1^3D_1$ state. 

No level splitting suggesting $\psi-X$ mixing seems to be at work here for in that case we would have expected the $\psi$ to be lighter than the $1^3D_1$ level~\footnote{We have tried to learn from the general structure of the levels  of exotic resonances using methods mediated by Random Matrix Theory~\cite{Cirillo:2011ia}. The approach has interesting potentialities on discerning between the level repulsion situation expected for a multiquark system and the Poisson distribution of levels which should  characterize a charmonium system. Its limitation resides basically on the scarcity of data available.}.

\begin{figure}[t]
\begin{center}
\epsfig{file=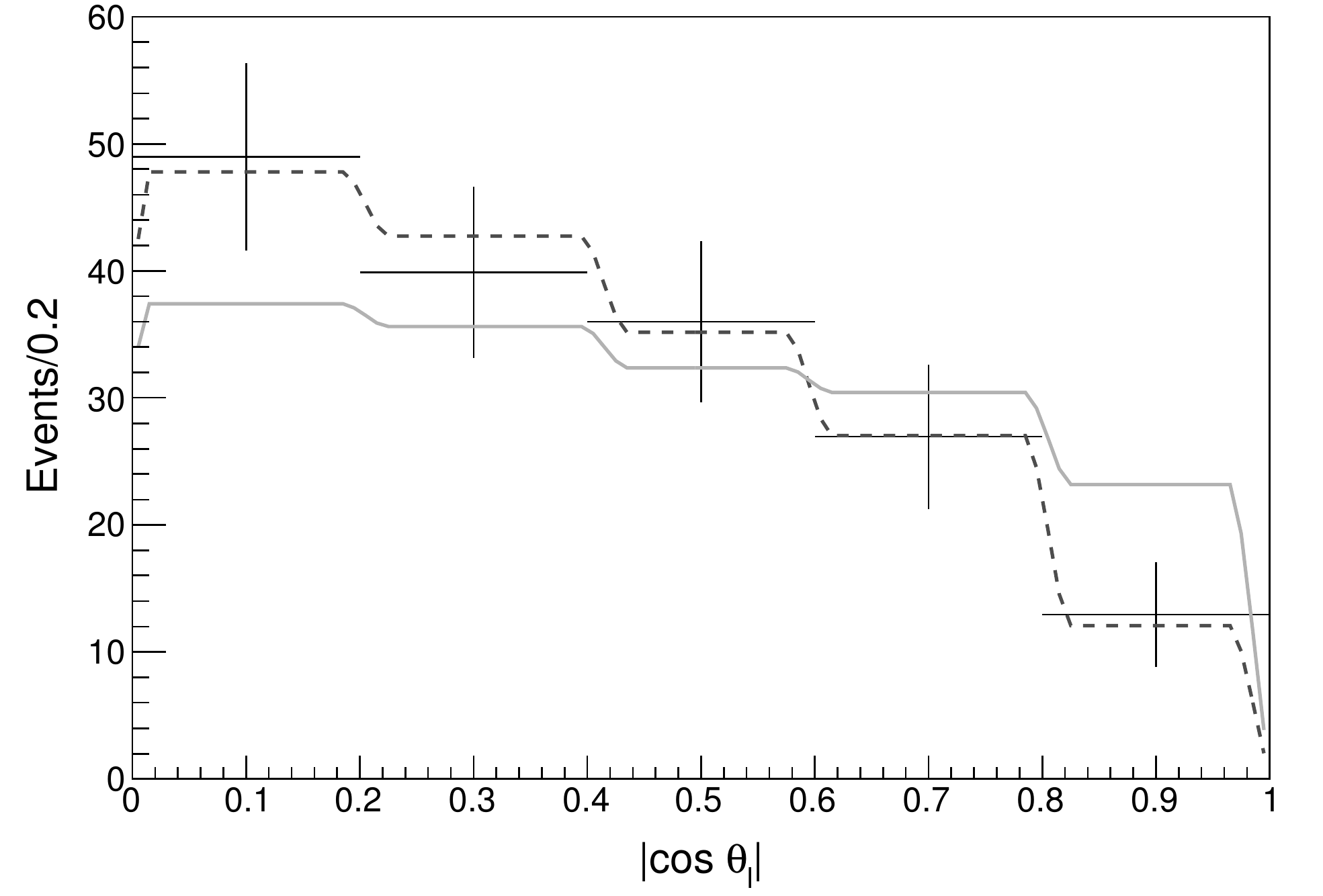,height=4cm}
\epsfig{file=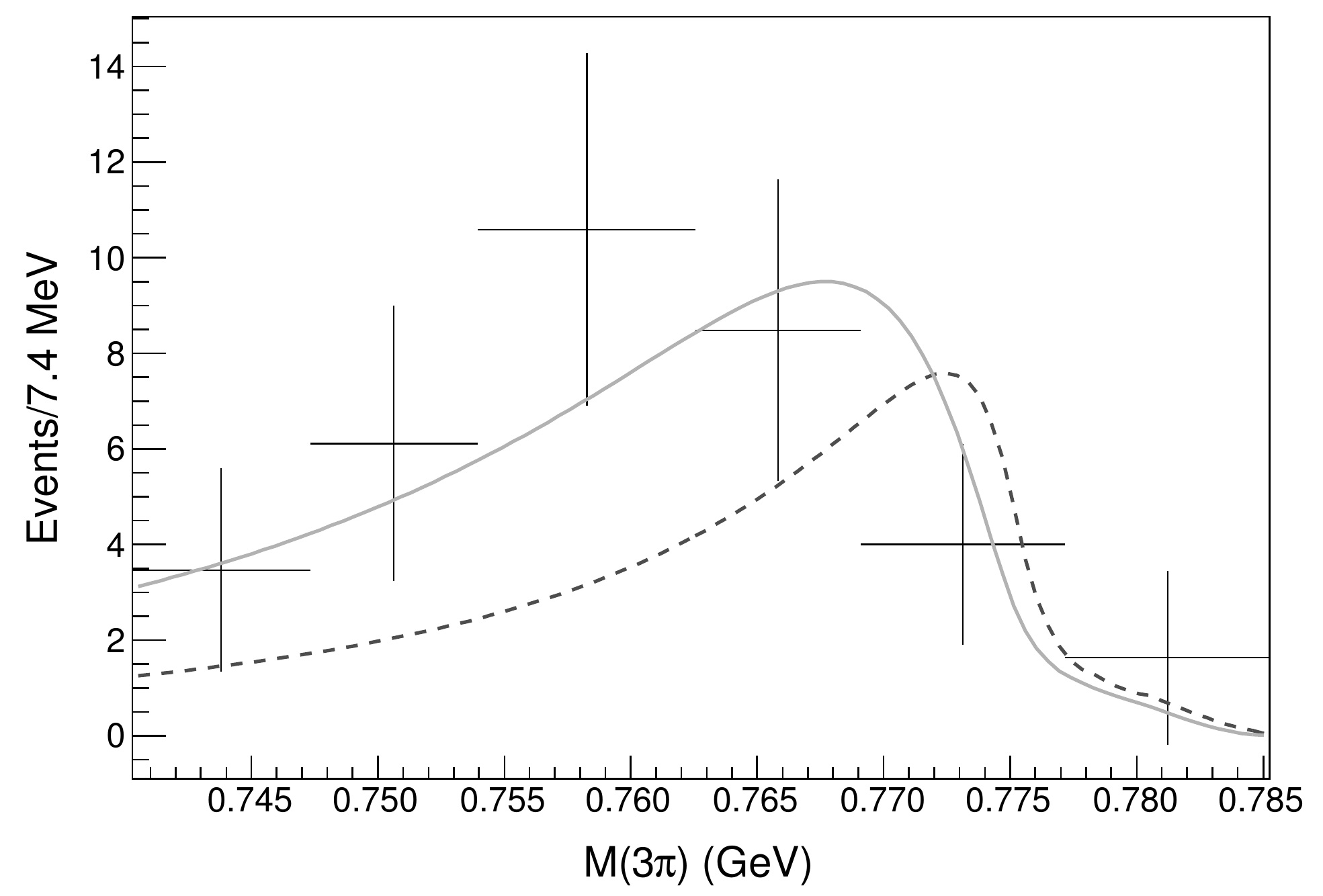,height=4.1cm}
\caption{\it Distribution of $ |cos\theta_l |$ (as defined in Ref.~\protect\cite{Choi:2011fc}) in $X(3872)\to J/\psi\pi^+\pi^-$ decays (left)  and of the $\pi^+\pi^-\pi^0$ invariant mass in $B\to X(3872)K$, $X(3872)\to J/\psi\pi^+\pi^-\pi^0$ decays (right). The full (dashed) curve is the expected spectrum under the hypothesis $J^{PC}=2^{-+}$ $(1^{++})$, as obtained in Ref.~\protect\cite{Faccini:2012zv}. }
\label{fig:spin}
\end{center}
\end{figure}

As another drawback of the charmonium $1^1D_2$ interpretation of the $X$ we mention that, to our knowledge, there is only one computation of the gluon fragmentation into a $1^1D_2$ state~\cite{Cho:1994qp} and it fails to reproduce the observed prompt cross section. We have updated the original calculation  with the latest PDF sets  finding that the production cross section expected at CDF is $\sigma(p\bar p\to 1^1D_2)\simeq 0.6$~nb, way smaller than the $\approx 30$~nb measured by CDF for $X$ prompt production. 

Because of these reasons it is really urgent to ascertain the $X$ quantum numbers. The most recent related measurements are the invariant mass and angular distributions published by Belle~\cite{Choi:2011fc} and BaBar~\cite{delAmoSanchez:2010jr} on $B\to X K$, $X\to J/\psi\pi\pi\pi^0$ and $X\to J/\psi\pi\pi$ respectively. A reanalysis of the invariant mass distributions was carried out with a Blatt-Wesskopf approach~\cite{Hanhart:2011tn}.

Very different conclusions came instead from a recent reanalysis of both angular and mass distributions~\cite{Faccini:2012zv}. In this paper the amplitudes were computed using a general relativistic formalism  to include in the calculations, especially when making angular analyses, all the spin correlations. In addition, a statistical approach that properly accounts for the fact that the distributions have different discrimination power between the spin hypotheses has been implemented.

The fit results are shown in Fig.~\ref{fig:spin} for the two most sensitive distributions. They lead to the conclusion that, considering the $X\to J/\psi\pi\pi$  sample alone, the hypothesis that the quantum numbers of $X$ are $J^{PC}=2^{-+}$ is excluded at 99.9\% C.L., while the $J^{PC}=1^{++}$ hypothesis is consistent with data. Conversely, considering the $X\to J/\psi\pi\pi\pi^0$ sample alone, the $J^{PC}=1^{++}$ hypothesis is excluded at 99.9\% C.L., while the $J^{PC}=2^{-+}$  is consistent with data.

This topic remains therefore open to very interesting developments.

\subsection{Tetraquarks}
The symmetry approach to the problem of $X,Y,Z$ spectroscopy heads for  multiquark interpretations. It is convenient to use diquarks as building blocks in order to reduce the number of possible states. Heavy-light diquark-antidiquark particles~\cite{Maiani:2004vq} fill up octets in the very same way as $q\bar q$ combinations do.  This in turn implies the prediction of a number of states. For example almost degenerate charged $X$ particles should exist forming an isotriplet $X^{+},X^{0},X^{-}$. 
There is also the  possibility that $X^{\pm}$ could be unobservable having  very broad widths~\cite{Terasaki:2011ma} and, indeed, have not been observed.

The tetraquark model also gives rather straightforward indications about what to expect in the beauty sector, predicting a replica of the charm spectroscopy  with hidden beauty. Interesting considerations on this topic can be found in Refs.~\cite{Ali:2009es,Ali:2011vy,Ali:2010pq,Ali:2011ug} and in Ref.~\cite{Ebert:2008se}.

The proliferation of states is  also typical of the molecule picture. We should expect a number of loosely bound molecules. They should be made up with very narrow mesons, narrower that the binding energy of the state. In the case of the $X(3872)$, if a total width larger than the total $D^{*}$ is found, the molecule model is excluded. Narrow open charm mesons $D,D^{*},D_{s},D_{s}^{*},D_{s0},D_{s1}^{*},D_{s1}^{\prime}$ with $q=u,d,s$ allow for 21 possible (loosely bound) molecules which should be observed. 

In the case of the $X(3872)$, the observation of isospin violations as shown by a ratio of branching ratios ${\cal B}(X\to J/\psi\; \rho)/{\cal B} (X\to J/\psi\; \omega)\simeq 1$, requires, in the tetraquark model, to have  two, instead of one, neutral $X$'s,
The second neutral $X$ is expected to be at an `hyperfine' separation from $X(3872)$, between $5$ and 10~MeV. 

The two neutral $X$ are  $X_{u}=[cu][\bar c\bar u]$ and $X_{d}=[cd][\bar c\bar d]$, {\it i.e.}, two isospin impure states each containing both $I=0,1$. 
Actually the $u\bar u$ pair in the $X_{u}$ tetraquark could convert into $d\bar d$ with an amplitude weighted by $\delta$ and this would modify the mass matrix in the direction of  aligning it along the isospin basis
\be
M=
\begin{pmatrix}
2m_{u}+2m_{c}&0\\
0&2m_{d}+2m_{c}\\
\end{pmatrix}
+\delta
\begin{pmatrix}
1&1\\
1&1\\
\end{pmatrix}
\ee
In order to have $X_{u}$ and $X_{d}$, {\it i.e.}, to be aligned in the flavor basis, we have to assume that $\delta$ is small at the mass scale $m_{c}$, an assumption which is particularly delicate as 
$\delta$ might depend more effectively on $\Lambda_{QCD}$ rather than on the heavy quark mass.
But what if we were actually observing two different states: a $2^{-+}$ and a $1^{++}$ with two different decay modes --- and no isospin violations at all? Can this be compatible with data?

\subsection{Tetraquarks and molecules}
Both models have loopholes and fail in different ways to give a consistent picture coherently describing  the $X,Y,Z$ spectroscopy. The hadron molecule interpretation has been evolving with time from the initial `deuson' formulation~\cite{Tornqvist:2004qy}, to the various attempts to understand the $X(3872)$ as a $D\bar D^{*}$ bound state held together by a pion-exchange mechanism~\cite{swansonrept} (and refereneces therein), up until the most recent ideas on universal Feshbach-resonance/hadron-molecule near-threshold behavior~\footnote{See the slides  of E. Braaten's talk at Charm 2012}.   

In the following we qualitatively
illustrate how the tetraquark/molecule dicothomy might eventually be accidental.
Let us use a toy-QCD where $\Lambda_{QCD}$ is a tunable parameter.

The mass of the $X$, like that of any other hadron, has to change by varying the QCD parameters like light quark masses $m_{q}$, heavy quark masses $m_{Q}$ and $\alpha_{s}$, $\Lambda_{QCD}$. Let us discuss qualitatively the variability with $\Lambda_{QCD}$, or better $\Lambda_{QCD}/m_{q}$, while keeping $m_{Q}$ fixed at some physical scale $\mu$.

Assume $\Lambda\lll \Lambda_{QCD}=200$~MeV. In this case, at the mass  scale of ordinary baryons, say $\mu=m_{p}$, the strong coupling constant $\alpha_s$ would be so small to make the proton a non-relativistic loosely bound state of three quarks with $m_{q}\simeq m_{p}/3=300$~MeV. The pion would then be much heavier ($\sim 600$~MeV) and spin-spin interactions, capable of shifting its mass, would  sensibly be depressed by the high value of $m_q$. Inter-spin forces are  back at work, making the pion lighter and eventually the lightest of all hadrons, as soon as $\Lambda$ is tuned back towards the ordinary $\Lambda_{{QCD}}$ value as soon as  $\Lambda > m_{q}$. 

Nonetheless we might assume heavy-light mesons (such as those containing one charm quark) to almost retain their same mass as $\Lambda$ is varied from  low values up to the standard one. This means that the $D+\bar D^{*}$ mass threshold is almost independent on $\Lambda$: spin-spin interactions in heavy-light mesons are always off due to $1/m_{c}$ suppression.

On the contrary spin-spin interactions between the two light quarks in a compact $cq\bar c \bar q$ tetraquark  might be responsible for a sensible change in the tetraquark mass as, starting from low $\Lambda$ values, the region is approached where $\Lambda > m_{q}$. This means that in a diagram mass vs. $\Lambda$ there could be a crossing point reached from above between the varying mass of the tetraquark and the threshold mass value  $D +\bar D^{*}$. 
\begin{figure}[bht]
\begin{center}
\epsfig{file=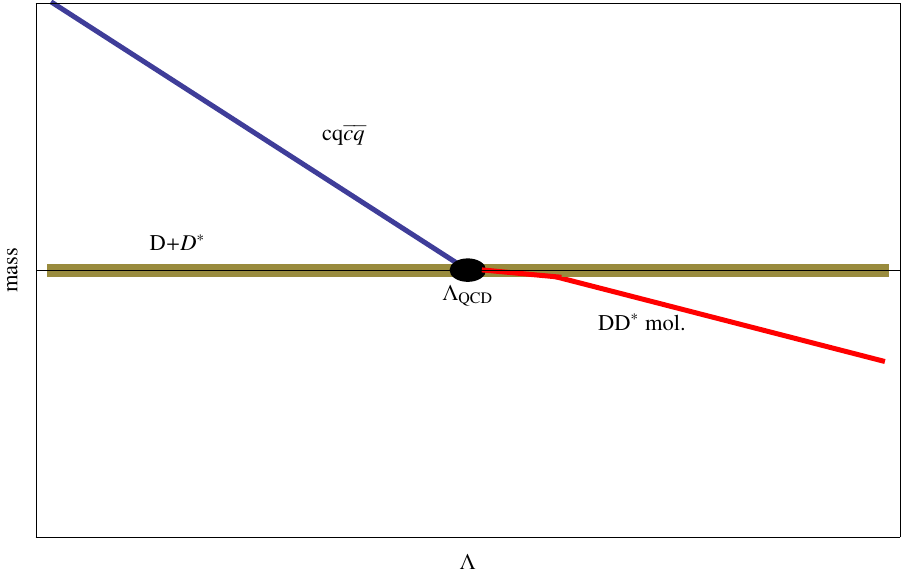,height=4.5cm}
\caption{\it Hypothetic plot on the dependency of the mass of a tetraquark and of $D\bar D^{*}$ bound state as $\Lambda$ is varied. Here $\Lambda_{1}=\Lambda_{2}=\Lambda_{QCD}$. The lower (red) curve, representing the mass of $DD^{*}$ as a function of $\Lambda$, reaches the threshold $D+D^{*}$ at $\Lambda_{QCD}$. For lower values of $\Lambda$ it cannot go through threshold for above that, the molecule falls apart. As for the blue curve, representing the mass of a tetraquark decreasing as $\Lambda$ increases, it could in principle go across the thresholds for $\Lambda>\Lambda_{QCD}$. $\Lambda\simeq\Lambda_{QCD}$, the physical value, might indeed be a special point where tetraquarks and molecules have a universal behavior.}
\label{scheme}
\end{center}
\end{figure}
Suppose that the crossing is at some $\Lambda=\Lambda_{1}$ (in Fig.~\ref{scheme} we actually assume $\Lambda_{1}\simeq \Lambda_{QCD}$). 

Suppose now that the Hamiltonian of strong interactions allows a deeply bound state molecule $D\bar D^{*}$ at some $\Lambda>\Lambda_{QCD}$. Decreasing $\Lambda$ towards $\Lambda_{QCD}$ (at some physical renormalization scale $\mu$) the binding energy of the molecule should decrease whereas the masses of the single meson components vary only very mildly. In addition, being  spin-spin interactions pointlike, they get less effective because of the larger size of  the less and less deeply bound molecule.  Assume that the mass of the molecule coincides with the threshold $D+D^{*}$ at some $\Lambda=\Lambda_{2}$. Above threshold the bound state fades away in the continuum spectrum.

The most interesting case would be the one in which $\Lambda_{1}\simeq \Lambda_{2}\simeq \Lambda_{QCD}$, {\it i.e.}, strong interaction dynamics should allow  a tetraquark state above threshold (where molecules with the same components of threshold are absent) for some $\Lambda<\Lambda_{QCD}$ and a deeply bound molecule for $\Lambda>\Lambda_{QCD}$: these two states coincide at $\Lambda_{QCD}$.  

Of course experiments are performed only at one point: $\Lambda_{QCD}$. Lattice simulations could be very instructive to explore this possibility, being able to simulate QCD with a pion mass ranging from $140$ to $500$~MeV. In this respects, close to threshold resonances could be seen as interwind combinations of tetraquarks and molecules.

\section{Conclusions}
We have briefly commented on some ideas animating the debate on $X,Y,Z$ spectroscopy and especially we have listed some of the experimental targets to aim at. We have focused especially  on the problem of the $X$ spin, which could subtend some new understanding of what the $X$ actually is.  

The $2^{-+}$ option for the $X(3872)$ is so much at odds with main interpretations (molecule, tetraquark, charmonium) to be considered almost as ``heresy''. We have reasons to believe that the contradictory information which seems to emerge from data available could  offer some unexpected understanding of the $X$ and  its possible close partners.

As shown in this paper, there is an entire program of measurements which should be carried out to help in recognizing a pattern in this sector of charmonium spectroscopy. In several cases data are available and simply have not been analyzed. The aim is not that of elaborating a taxonomy of states through the accumulation of  data. More information is necessary to understand if there is something new to learn about the dynamics of strong interactions at large distance, molecules, or, at low distances (with respect to the range of strong interactions), diquarks-antidiquarks or other aggregations of quark/gluon matter. We have indications for  thinking that this is indeed the case, and this  provides the motivation to continue investing  effort in  this research field.
 
 \bibliography{MPLA2012} 
 \bibliographystyle{JHEP}                                                                                             

\end{document}